\documentclass[twocolumn,aps,superscriptaddress,nofootinbib]{revtex4}

\usepackage{amsmath,amssymb}
\usepackage{graphicx}
\usepackage{dcolumn}
\usepackage{bm}
\usepackage[usenames]{color}
\usepackage[breaklinks=true,colorlinks=true,linkcolor=blue,urlcolor=blue,citecolor=blue]{hyperref}
\usepackage{ulem}
\usepackage{amsmath}
\usepackage{float}
\usepackage{ulem}

\begin{document}

\title{Polymer translocation in an environment of active rods}
 
\author{Hamidreza Khalilian}
\email{khalilian@ipm.ir}
\affiliation{School of Nano Science, Institute for Research in Fundamental Sciences (IPM), 
19395-5531, Tehran, Iran.}

\author{Jalal Sarabadani}
\email{jalal@ipm.ir}
\affiliation{School of Nano Science, Institute for Research in Fundamental Sciences (IPM), 
19395-5531, Tehran, Iran.}

\author{Tapio Ala-Nissila}
\affiliation{Department of Applied Physics and QTF Center of Excellence, Aalto University, P.O. Box 11000, FI-00076 Aalto, Espoo, Finland.}
\affiliation{Interdisciplinary Centre for Mathematical Modelling and Department of Mathematical Sciences, 
Loughborough University, Loughborough, Leicestershire LE11 3TU, UK.}

\begin{abstract}
We consider the dynamics of a translocation process of a flexible linear polymer through a nanopore into an environment of active rods in the {\it trans} side. Using Langevin dynamics simulations we find that the rods facilitate translocation to the {\it trans} side even when there are initially more monomers on the {\it cis} than on the {\it trans} side. Structural analysis of the translocating polymer reveals that active rods induce a folded structure to the {\it trans}-side subchain in the case of successful translocation events. By keeping the initial number of monomers on the {\it cis}-side subchain fixed, we map out a state diagram for successful events as a function of the rod number density for a variety of system parameters. This reveals competition between facilitation by the rods at low densities and crowding that hinders translocation at higher densities.  
\end{abstract}

\maketitle

\section{Introduction} \label{intro}

Polymer translocation through a synthetic or biological nanopore has become an active field of research since the seminal experimental works by Bezrukov {\it et al.} in 1994 \cite{Bezrukov} and two years later by Kasianowicz {\it et al.} \cite{KasiPNAS1996}, and by Sung and Park \cite{SungPRL1996} in 1996 on the theoretical side \cite{Muthukumar_book,Tapio_review,MilchevJPCM2011,jalalJPCM2018,mellerPRL2001,Smith_Nature_2001,%
Storm2003,BrantonPRL2003,StormNanoLett2005,Keyser_2006,%
Keyser_2009,Bulushev_2015,MuthukumarJCP1999,%
Kantor_PRE2004,GrosbergPRL2006,aksimentievNanolett2008,rowghanian2011,%
SakauePRE2007,SakauePRE2010,SakauePRE2012,TapioPRE2007,%
ikonen2012a,ikonen2012b,%
jalalJCP2014,jalalJCP2015,JalalEPL2017,jalalSciRep2017,slaterJCP2012,slaterPRE2010,slaterPRE2009,%
hamidJCP2013,%
jalalPolymers2018,jalalPolymers2019,jalalJPCM2020_1,jalalJPCM2020_2,golestanianPRL2011,%
menais_2017,menais_2018,chou_nanoletter_2012,Aniket_2015,Aniket_2020,Fazli_arxiv,Jalal_Ralf_Tapio_PRR_2022}.
The translocation process has many applications in different areas, such as transport of mRNA through nuclear membrane pores~\cite{nuclear}, drug delivery and DNA sequencing~\cite{drug_DNA,Branton_Naturebio,Deamer_Naturebio} and gene transfer between bacteria~\cite{bact}. 
Biological polymers often translocate through a nanopore embedded in a membrane into an environment composed of other biological organisms~\cite{Alberts}, such as diffusive or fixed spherical obstacles \cite{gopinathan,chen,yu,samadi}, chaperones \cite{abdolvahab,emamyari,xu} and other biomolecules or even micro-organisms. 
\begin{figure}[b]
	 \hspace{-0.5cm}
	 \begin{center}
     \includegraphics[width=0.8\columnwidth]{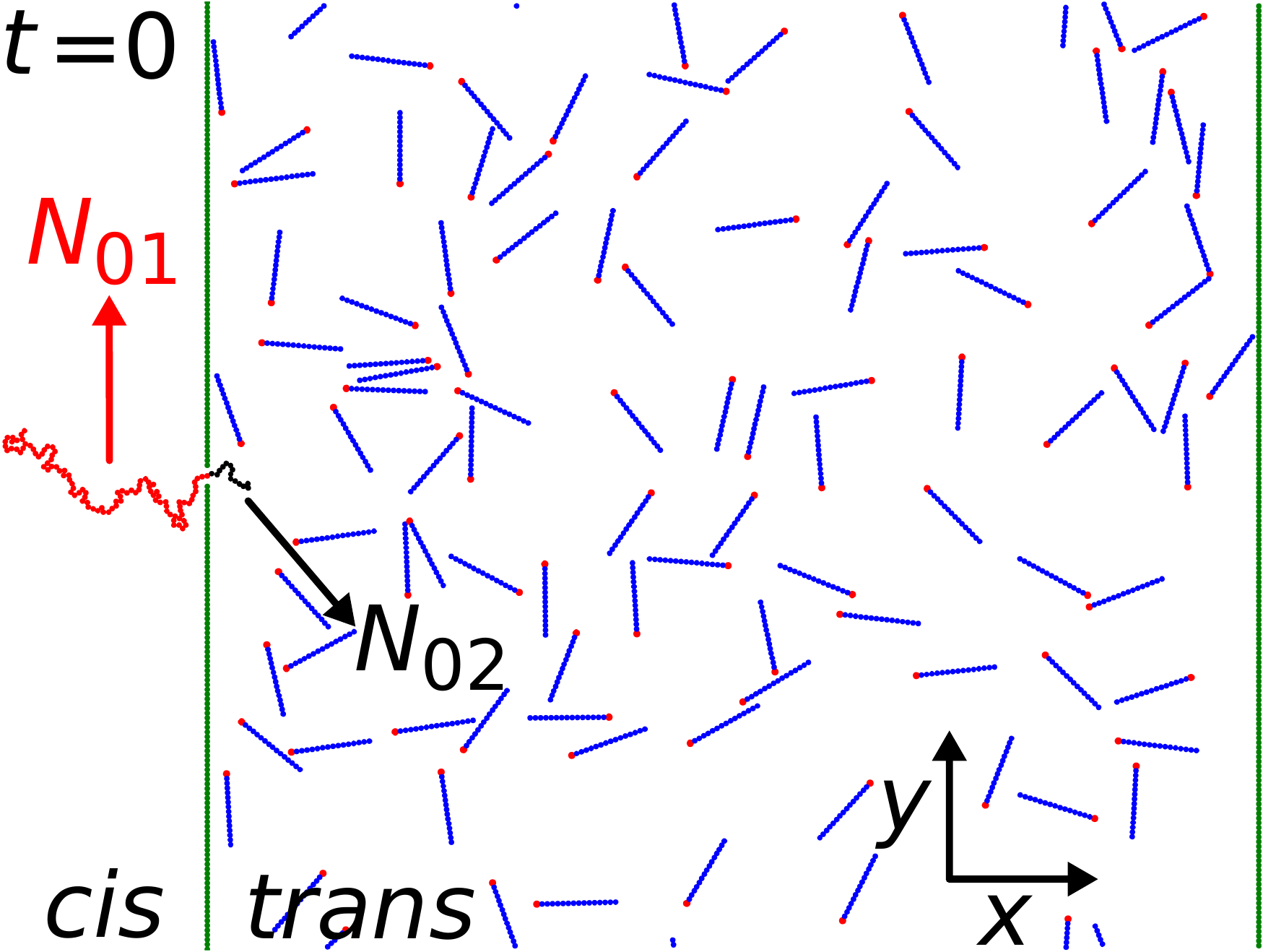}
     \end{center}
     \caption{Configuration of the system after equilibration and before the actual translocation process. $N_{01}$ and $N_{02}$ are the initial contour lengths of the {\it cis}- and {\it trans}-side polymer subchains, respectively. After equilibration the self-propelling (SP) force $F_{\textrm{sp}}$ is switched on and acts on the head of each rod (filled red circles in the {\it trans} side) from its tail to its head. In all simulations, the initial {\it cis}-side subchain contour length is fixed at $N_{01} = 100$. There is one monomer at the pore and thus the total contour length of the polymer is $N_{0}=N_{01}+N_{02}+1$. The directions perpendicular and parallel to the membrane (green vertical line) are denoted by $x$ and $y$, respectively. 
} \label{model}
\end{figure}

The nature of the dynamics of polymer translocation depends on the details of the setup. There are multiple different scenarios including the {\it pore-driven} case~\cite{SakauePRE2007,rowghanian2011,ikonen2012a,jalalJCP2014}, wherein the external driving force only acts on those monomers inside the nanopore, the {\it end-pulled} case~\cite{TapioPRE2007,JalalEPL2017}, in which the external driving force acts on the head monomer of the polymer either by atomic force microscope~\cite{AFM_Ritort} or by optical or magnetic tweezers~\cite{Keyser_2006,Smith,Sischka,Bulushev}, and the {\it unbiased} case with no external drive~\cite{slaterPRE2010,slaterJCP2012,slaterPRE2009}. 
In biological systems chaperons and other active agents can strongly influence translocation processes. To this end, pore-driven polymer translocation into an environment of active spherical particles has been studied and it has been shown numerically that the crowding effect in two dimensions increases the translocation speed provided that the particles activity is high enough \cite{pu}. Also, using the same model, the translocation time has been obtained as a function of the size of spherical active particles, and the most favorable case occurs when the size of the active particles is comparable to the segment (monomer) size of the polymer \cite{tan}.

In the externally unbiased translocation process there must be an effective driving force on the chain to overcome the repulsive entropic force. Without an external drive, an imbalance may come from a net attractive entropic force due to unequal fluctuations of the spatial configurations of the polymer between the {\it cis} and the {\it trans} sides  \cite{slaterJCP2012,slaterPRE2010,slaterPRE2009}. For example, the presence of chaperons~\cite{abdolvahab}, attractive binding particles~\cite{Aniket_2015} or active particles~\cite{hamidPRR2021} in the {\it trans} side may lead to a directed motion of the polymer either to the {\it cis} or to the {\it trans} side depending on the values of the system parameters~\cite{Aniket_2015,pu,hamidPRR2021}. In particular, we have recently shown in Ref.~\cite{hamidPRR2021} that active rodlike particles in the {\it trans} side can 
induce an effective driving force which facilitates translocation when part of the chain is initially in the {\it trans} side. Using extensive computer simulations and the iso-flux tension propagation (IFTP) theory we obtained the scaling form of the mean translocation time as a function of the initial polymer contour length in the {\it cis} side, the rod activity,  and the rod length \cite{hamidPRR2021}.

In Ref.~\cite{hamidPRR2021} the initial contour length of the chain in the {\it trans} side was fixed to a constant value of $N_{02}=100$ monomers while the other relevant parameters were varied such that translocation events were successful. An interesting question then remains concerning the role of $N_{02}$ in determining successful translocation events. To this end, in the present work we use extensive Langevin dynamics simulations to investigate this issue.
While the initial contour length of the {\it cis}-side subchain is fixed at $N_{01} = 100$, the contour length of the polymer subchain in the {\it trans} side $N_{02}$ is varied together with other system parameters. This allows us to map out a "state diagram" with a boundary separating successful and failed translocation events.

A particularly interesting finding in our work is that due to their shape anisotropy the rods may align with the membrane and induce {folds} on the {\it trans}-side polymer subchain. This facilitates directed translocation towards the {\it trans} side for cases where $N_{02} < N_{01}$, which would not be possible without the active rods. The state diagram also reveals that for fixed $N_{01}$, the boundary between successful and unsuccessful events has non-monotonic dependence on the number density of the rods due to rod crowding, which will eventually hinder translocation for higher densities.

The paper is structured as follows. The Langevin dynamics (LD) simulation model is explained in detail in Sec.~\ref{SIM}. Section~\ref{RES} is devoted to the numerical results. Finally, our summary and conclusions can be found in Sec.~\ref{CONCLUSION}.

\section{simulation model} \label{SIM}

We consider a two-dimensional simulation box of size $L_x = 400 \sigma$ and $L_y = L_x /2$ in the $x$ and $y$ directions, respectively (see Fig.~\ref{model}). A one-dimensional membrane composed of particles with size $\sigma$ separated by $\sigma$ from each other divides the simulation box in the $y$ direction. A nanopore of radius $1.5 \sigma$ is located at $x=y=0$ in the membrane. A flexible linear polymer with a total number of monomers $N_0$  modeled by the bead-spring model \cite{grest} is set in the simulation box such that one of its monomers is inside the nanopore. The {\it cis}- and {\it trans}-side subchains contain $N_{01}$ and $N_{02}$ monomers, respectively, and $N_{0}=N_{01}+N_{02}+1$. In the {\it trans} side, there are $N_{\textrm{r}}$ rigid rods, each composed of $L_{\textrm{r}}$ beads with diameter $\sigma$. Two walls (constructed similar to the membrane) parallel to the $y$ direction are located at $x=-200 \sigma$ and $x=200 \sigma$ and periodic boundary conditions are applied in the $y$ direction. 

The bonded interaction between consecutive monomers in the polymer is a combination of a finitely extensible nonlinear elastic (FENE) potential
\begin{equation}
U_{\textrm{FENE}}(r) = -\frac{1}{2}k R^{2}_{0}\ln(1-r^2 / R_0^2),
\label{fene}
\end{equation}
where $k$, $R_0$ and $r$ are the spring constant, the maximum allowed distance, and the distance between the consecutive monomers, respectively, and the Weeks-Chandler-Anderson (WCA) potential
\begin{equation}
 U_{\textrm{WCA}}(r) = \left\lbrace
  \begin{array}{l l}
	U_{\textrm{LJ}} (r) - U_{\textrm{LJ}} (r_{\textrm{c}}), 
     & \text{$r < r_{\textrm{c}}$};\\
    0, & \text{$r \geq r_{\textrm{c}}$},
  \end{array}
\right. 
\label{slj}
\end{equation}
where $r_{\rm c}=2^{1/6}\sigma$ is the cut-off radius. The term  $U_{\textrm{LJ}}(r)$ is the Lennard-Jones (LJ) potential
\begin{equation}
 U_{\textrm{LJ}}(r) =  4\varepsilon \bigg[ \bigg( \frac{\sigma}{r} \bigg)^{12} - \bigg( \frac{\sigma}{r} \bigg)^{6} \bigg],
\label{lj}
\end{equation}
with $\varepsilon$ and $r$ as the potential well depth and the distance between two given particles, respectively.
The non-bonded monomers as well as the monomers and the wall particles, and also membrane particles and the rod beads, interact each other via the WCA excluded volume interaction.

The LD equation is employed to describe the dynamics of the position of the $i^{\rm th}$ monomer of the polymer as
\begin{equation}
{M} \ddot{\vec{r}}_{i} = -\eta \dot{\vec{r}}_{i} - \vec{\nabla} U_{\textrm{m}i} + \vec{\xi}_{i}(t),
\label{ldp}
\end{equation}
where $\eta$ and $U_{\textrm{m}i}$ are the solvent friction coefficient and the sum of all interactions for the $i^{\rm th}$ monomer, respectively. $\vec{\xi}_i$ is the white noise with $\langle \vec{\xi}_{i} (t) \rangle  = \vec{0}$ and $\langle \vec{\xi}_{i} (t) \cdot \vec{\xi}_{j} (t') \rangle  = 4 \eta k_{\textrm{B}} T \delta_{ij} \delta (t - t')$, where $k_{\textrm{B}}$, $T$, $\delta_{ij}$ and $\delta (t -t')$ are the Boltzmann constant, the temperature, the Kronecker and Dirac delta functions, respectively. 

The time evolution of the $i^{\rm th}$ bead of each rod is governed by the following Langevin equation:
\begin{equation}
{M} \ddot{\vec{r}}_{i} = -\eta \dot{\vec{r}}_{i} + F_{\textrm{SP}} \delta_{i\textrm{h}} \hat{{e}} - \vec{\nabla} U_{\textrm{r}i} + \vec{\xi}_{i}(t),
\label{ldr}
\end{equation}
where $F_{\textrm{SP}}$ and $\textrm{h}$ stand for the self-propelling (SP) force and the head bead of the rod, respectively, $\hat{{e}}$ is the unit vector parallel to the rod's main axis from the tail to the head bead, and $U_{\textrm{r}i}$ is the sum of all interactions on the $i^{\rm th}$ bead of the rod. 

$M, \sigma$ and $\varepsilon$ are employed as the units for mass, length and energy, respectively. In our simulations, we use $M=1$ as mass of each monomer in the polymer and bead in the rod, and set $\sigma = 1$ and $\varepsilon=1$. The temperature is kept at $k_{\textrm{B}}T=1.2$ and the solvent friction coefficient is $\eta=0.7$. The integration time step is $\textrm{d} t=0.001 \tau_{0}$ where $\tau_{0}=\sqrt{M\sigma^{2}/\varepsilon}$ is the simulation time unit. The spring constant is set to $k=30$, and $R_{0}=1.5$. LAMMPS package \cite{lammps} has been used to perform the simulations and the results have been averaged over 2000 uncorrelated trajectories. 

In our model each bead size corresponds to the single-stranded DNA Kuhn length which is about $1.5$nm, and the interaction strength at room temperature $T=295$~\!K is $3.39 \times 10^{-21}$~\!J. Thus, the timescales and force in LJ units are approximately $32.1$~\!ps and $2.3$~\!pN, respectively. 

Before starting the actual translocation process, when the monomer inside the pore is fixed such that there are $N_{01}$ and $N_{02}$ monomers in the {\it cis} and {\it trans} sides of the simulation box, respectively, and in the absent of the SP force, the system is equilibrated for $t_{\textrm{eq}}= 5 \times 10^{4}\tau_{0}$. Then, after equilibration, the polymer is released and the SP force $F_{\textrm{sp}}$ is simultaneously switched on for all the rods.
The number density of the active rods (ARs) is given by $\rho_{\textrm{r}} = L_{\textrm{r}} N_{\textrm{r}} / ( L_x L_y /2 )$ and it is low enough such that after equilibration the rods are in a spatially-uniform isotropic phase.

For all simulations, the initial length of the polymer subchain in the {\it cis} side is fixed at $N_{01}=100$, and at fixed values of $F_{\textrm{sp}}$ and $\rho_{r}$ the value of $N_{02}$ (initial length of the polymer subchain in the {\it trans} side) has been varied to find its minimum value for which more than 50$\%$ of the translocation events to the {\it trans} side are successful. The LD simulations have been performed for the SP force $F_{\textrm{sp}}=32$ with the rod lengths $L_{\textrm{r}}=4, 8, 16$ and $20$, and also for the rod length $L_{\textrm{r}}=16$ with SP forces $F_{\textrm{sp}}=4, 8, 16$ and $32$, including $N_{\textrm{r}}=1, 10, 20, 50, 100, 200$ and $320$ ARs in the {\it trans} side of the simulation box.

\section{Results} \label{RES}

Figure~\ref{model} shows a typical snapshot of the system just before the translocation process at $t=0$ in equilibrium state with the rods in the isotropic phase. When the SP force is switched on, the combination of alignment of the ARs with the membrane as well as interaction between ARs and the {\it trans}-side polymer subchain induces a net effective force towards the {\it trans} side \cite{hamidPRR2021}. Therefore, a tension front propagates along the backbone of the {\it cis}-side subchain while the polymer is sucked by the environment of ARs into the {\it trans} side. This mechanism allows the polymer to overcome the entropic barrier and facilitates the translocation process provided that the initial contour length of the polymer subchain in the {\it trans} side is long enough ({\it e.g.}, $N_{02}=100$ in our previous work~\cite{hamidPRR2021}). 

\begin{figure*}[t]
	\begin{minipage}{1.0\textwidth}
    \begin{center}
        \includegraphics[width=0.98\textwidth]{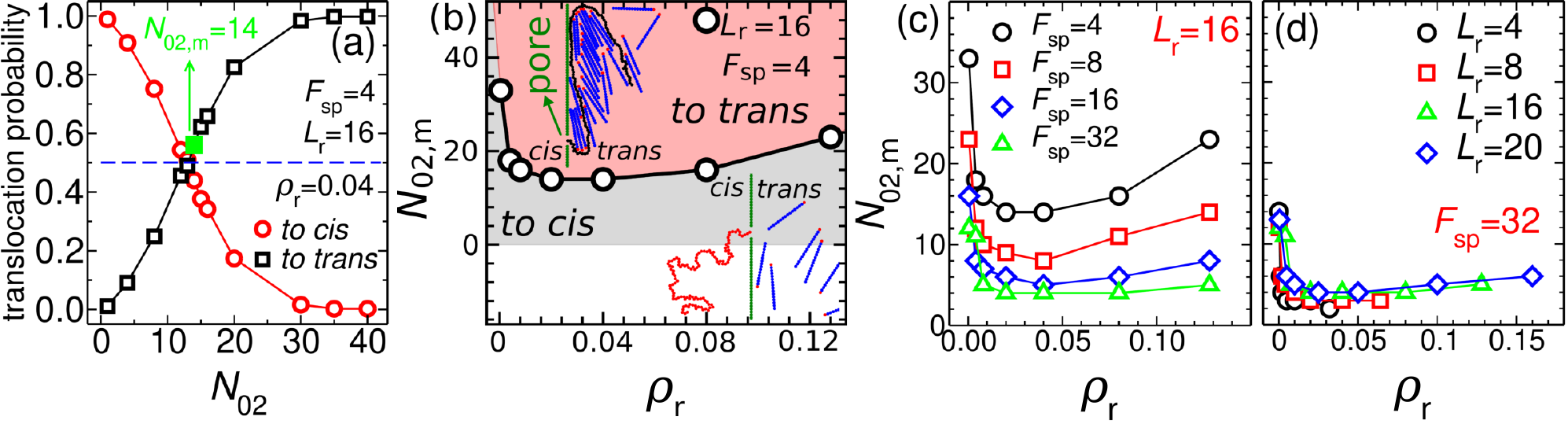}
    \end{center}
    \end{minipage}
\caption{(a) The probability of polymer translocation through a nanopore back to the {\it cis} side (open red circles) and into the {\it trans} side (open black squares) as a function of the initial value of the polymer subchain contour length in the {\it trans} side, $N_{02}$, for fixed values of $L_{\textrm{r}}=16$, $F_{\textrm{sp}}=4$ and $\rho_{\textrm{r}}=0.04$ ($N_{\textrm{r}}=100$). The filled green square at $N_{02,m}=14$ represents the boundary between the {\it to-cis} and {\it to-trans} states. 
The horizontal blue dashed-line is plotted at 1/2.
(b) A diagram showing the regions of {\it to-cis} and {\it to-trans} states in the $N_{\textrm{02,\rm m}} - \rho_{\textrm{r}}$ plane for fixed values of $L_{\textrm{r}}=16$ and $F_{\textrm{sp}}=4$. The open black circles connected by the solid line denote the boundary between the two regions. The top and the bottom insets show snapshots of final configurations of the system in the vicinity of the nanopore for a fixed value of $N_{02,\textrm{m}}=14$.
(c) Additional diagrams in the $N_{\textrm{02,m}} - \rho_{\textrm{r}}$ plane separating the  {\it to-trans} (above the curves) from the {\it to-cis} (below the curves) events for $L_{\textrm{r}}=16$ and various values of  $F_{\textrm{sp}} = 4$ (black line - circles), $8$ (red line - squares), $16$ (blue line - diamonds) and $32$ (green line - triangles). (d) The same as in panel (c) but for fixed $F_{\textrm{sp}} = 32$ and different values of $L_{\textrm{r}}= 4$ (black line - circles), $8$ (red line - squares), $16$ (green line - triangles) and $20$ (blue line - diamonds).
} 
\label{probability}
\end{figure*}

In the present work we focus on the influence of  the initial contour length of the {\it trans}-side polymer subchain on the dynamics of the translocation process. To this end, we keep the  initial contour length of the {\it cis}-side polymer subchain constant at $N_{01}=100$. 
%
When 50 \% or more of the translocation attempts are successful, we label the corresponding parameter sets as 
{\it to-trans} states while the unsuccessful events are labeled as {\it to-cis} states.

\subsection{Dependence of translocation events on system parameters} \label{states}

In order to distinguish between the {\it to-trans} and {\it to-cis} states, the values of the parameters such as $N_{\textrm{r}}, L_{\textrm{r}}$ and $F_{\textrm{sp}}$ are kept fixed and the length of polymer subchain in the {\it trans} side $N_{02}$ is varied. By finding the minimum value of $N_{02}$ that is labeled $N_{02,\textrm{m}}$ at which the translocation is {\it to-trans}, we can draw a diagram on the $N_{02,\textrm{m}} - \rho_{\textrm{r}}$ plane for fixed values of $L_{\textrm{r}}$ and $F_{\textrm{sp}}$.

Figure~\!\ref{probability}(a) shows the probability of polymer translocation into or out of the active environment in the {\it trans} side as a function of the initial value of the {\it trans}-side polymer subchain contour length $N_{02}$ for fixed values of $L_{\textrm{r}}=16, F_{\textrm{sp}}=4$ and {$\rho_{\textrm{r}} = 0.04$ corresponding to} $N_{\textrm{r}}=100$. By increasing the value of $N_{02}$ the probability of translocation into the {\it trans} side increases (black squares) while the probability of failed events decreases (red circles). The filled green square denoted by $N_{02,\rm m}=14$ identifies the boundary wherein for $N_{02}< N_{02,\rm m}$ the translocation event is {\it to-cis} state, and for $N_{02} \geq N_{02,\rm m}$ the state is {\it to-trans}.

Panel (b) in Fig.~\!\ref{probability} shows the state diagram in the $N_{02,\rm m}-\rho_{\textrm{r}}$ plane for fixed values of $L_{\textrm{r}}=16$ and $F_{\textrm{sp}}=4$. The value of $N_{\textrm{r}}$ has been varied and the diagram obtained from the translocation probabilities explained in Fig.~\ref{probability}(a). The pink region above and the gray below the black curve correspond to the {\it to-trans} and {\it to-cis} states, respectively. 
The top and the bottom insets show typical final configurations of the system with fixed values of  $L_{\textrm{r}}=16$, $F_{\textrm{sp}}=4$, $N_{02}=N_{02,\rm m}=14$ and $\rho_{\textrm{r}} = 0.04$ ($N_{\textrm{r}}=100$). As can be seen, the polymer is sucked into the active {\it trans}-side region if a {folded} structure is induced in the polymer configuration by the interaction between polymer and ARs (top inset). On the other hand, the polymer leaves the pore towards the {\it cis} side in the absence of such a {folded} structure (bottom inset). The configurations of the polymer in the {\it cis} and in the {\it trans} sides will be described in detail in subsection \ref{structure}.

To investigate the effect of SP force on the border between the {\it to-cis} and {\it to-trans} states, Fig.~\ref{probability}(c) shows a diagram in the $N_{02,\rm m} - \rho_{\textrm{r}}$ plane for fixed value of the rod length $L_{\textrm{r}}=16$ and various values of SP force $F_{\textrm{sp}}= 4$ (black line - circles), $8$ (red line - squares), $16$ (blue line - diamonds) and $32$ (green line - triangles). The regions above and below each curve represent the {\it to-trans} and {\it to-cis} states, respectively, for the corresponding SP force. Panel (d) in Fig.~\ref{probability} is the same as panel (c) but for fixed value of SP force $F_{\textrm{sp}}=32$ and different values of the rod length $L_{\textrm{r}}= 4$ (black line - circles), $8$ (red line - squares), $16$ (green line - triangles) and $20$ (blue line - diamonds). In the absence of ARs in the {\it trans} side, for $N_{01}=N_{02}=100$ the polymer translocates equally either to the {\it trans} side or retract back to the {\it cis} side, i.e. $N_{02}=N_{02,\rm m}=100$ (not shown here).

By increasing the number of ARs in the {\it trans} side for the values of the system parameters in the present study, the effective net force on the monomer inside the pore towards the {\it trans} side increases due to the interactions between the ARs and the {\it trans}-side polymer subchain. This increases the probability of successful events \cite{hamidPRR2021} leading to smaller values of $N_{02,\rm m}$. On the other hand, for large values of the rod length $L_{\textrm{r}}$ and increasing {the rod density $\rho_{\textrm{r}}$}, the crowding effect due to the presence of ARs in the vicinity of nanopore in the {\it trans} side can block translocation {\it to-trans}, which leads to larger values of $N_{02,\rm m}$. Moreover, for small values of the rod length, $N_{02,\rm m}$ becomes a monotonically decreasing function for increasing {$\rho_{\textrm{r}}$} due to negligible crowding.

\subsection{Mean translocation time} \label{ttime}

\begin{figure*}[t]\begin{center}
	\begin{minipage}{1.0\textwidth}
    \begin{center}
        \includegraphics[width=0.98\textwidth]{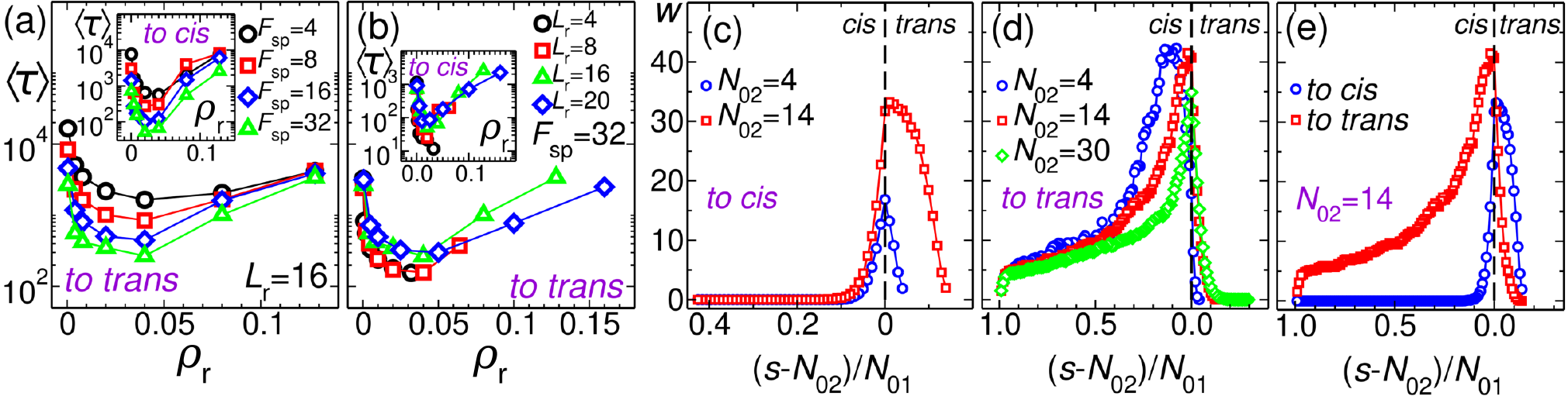}
    \end{center}
    \end{minipage}
\caption{
(a) Mean translocation time $\langle \tau \rangle$ as a function of the {rod density $\rho_{\textrm{r}}$} for fixed rod length $L_{\textrm{r}}=16$ and various values of the SP force  $F_{\textrm{sp}}$ in the {\it to-trans} state for the values of the {\it trans}-side polymer subchain contour length on the boundaries depicted in the Fig. \ref{probability}(c). Inset shows the same quantity as in the main panel for the {\it to-cis} state. (b) same as the panel (a) but for fixed value of the SP force $F_{\textrm{sp}}=32$ and various values of the rod length $L_{\textrm{r}}$. Inset presents the same quantity as the main panel  for the {\it to-cis} state.
(c) Waiting time distribution $w$ as a function of normalized and rescaled translocation coordinate $ ( s - N_{02} ) / N_{01} $ with $N_{02}=4$ (blue line - circles) and $14$ (red line - squares) for the {\it to-cis} state. (d) The same quantity as panel (c) but for the {\it to-trans} state with $N_{02}=4$ (blue line - circles), $14$ (red line - squares) and $30$ (green line - diamonds). (e) The same quantity as in panels (c) and (d) for fixed value of $N_{02}=14$ and for the {\it to-cis} (blue line - circles) and {\it to-trans} (red line - squares) states. In panels (c), (d) and (e) the values of the rod length, SP force and number of rods in the {\it trans} side are fixed at $L_{\textrm{r}}=16$, $F_{\textrm{sp}}=4$ and $\rho_{\textrm{r}} = 0.04$  ($N_{\textrm{r}}=100$), respectively, and the vertical black dashed line denotes the pore position.} 
\label{ttimes}
\end{center}
\end{figure*}

 The central quantity that reveals the dynamics of the translocation process at the global level is the mean translocation time which is the time it takes for the whole polymer to enter the {\it trans} side (translocation time for the {\it to-trans} state). On the other hand, the time it takes for the polymer to completely retract and enter the {\it cis} side will be defined as the translocation time for the {\it to-cis} state. 

In the main panel of Fig.~\ref{ttimes}(a) the mean translocation time $\langle \tau \rangle$ for the {\it to-trans} state is plotted as a function of the rod {density} in the {\it trans} side {$\rho_{\textrm{r}}$}, for fixed $L_{\textrm{r}} = 16$ and various values of the SP force $F_{\textrm{sp}} = 4$ (black line - circles), $8$ (red line - squares), $16$ (blue line - diamonds) and $32$ (green line - triangles) [cf. Fig.~\!\ref{probability}(c)]. The inset shows the same quantity as in the main panel but for the {\it to-cis} state. Panel (b) of Fig.~\!\ref{ttimes} is the same as panel (a) but for fixed values of the SP force $F_{\textrm{sp}} =32$ and for different values of the rod length $L_{\textrm{r}} = 4$ (black line - circles), $8$ (red line - squares), $16$ (green line - triangles) and $20$ (blue line - diamonds). Some typical translocation time distributions are shown in the SI.

As mentioned before, the top snapshot in the inset of Fig.~\!\ref{probability}(b) clearly shows that the polymer is sucked into the active environment in the {\it trans} side if a {folded} structure is induced on the {\it trans}-side polymer subchain by the ARs. The main panels in Figs.~\!\ref{ttimes}(a) and (b), which are for the {\it to-trans} state, show interesting non-monotonic behavior of the translocation time. For small densities the effective force first increases and the mean translocation time decreases together with $N_{02}$. The fastest translocation occurs around $\rho_{\rm r} \approx 0.04$ where also $N_{02,\rm m}$ has a minimum, after which the translocation time starts to increase as crowding of the rods becomes relevant. It is interesting to note that the position of the minimum is almost independent of the activity force and the rod length in the range of parameters studied here. In the insets of panels (a) and (b) in Fig.~\!\ref{ttimes} we show data for the {\it to-cis} state and the behavior of the translocation time is very similar to that of the {\it to-trans} state. 

\begin{figure*}[t]\begin{center}
    \begin{minipage}{1.0\textwidth}
    \begin{center}
        \includegraphics[width=0.98\textwidth]{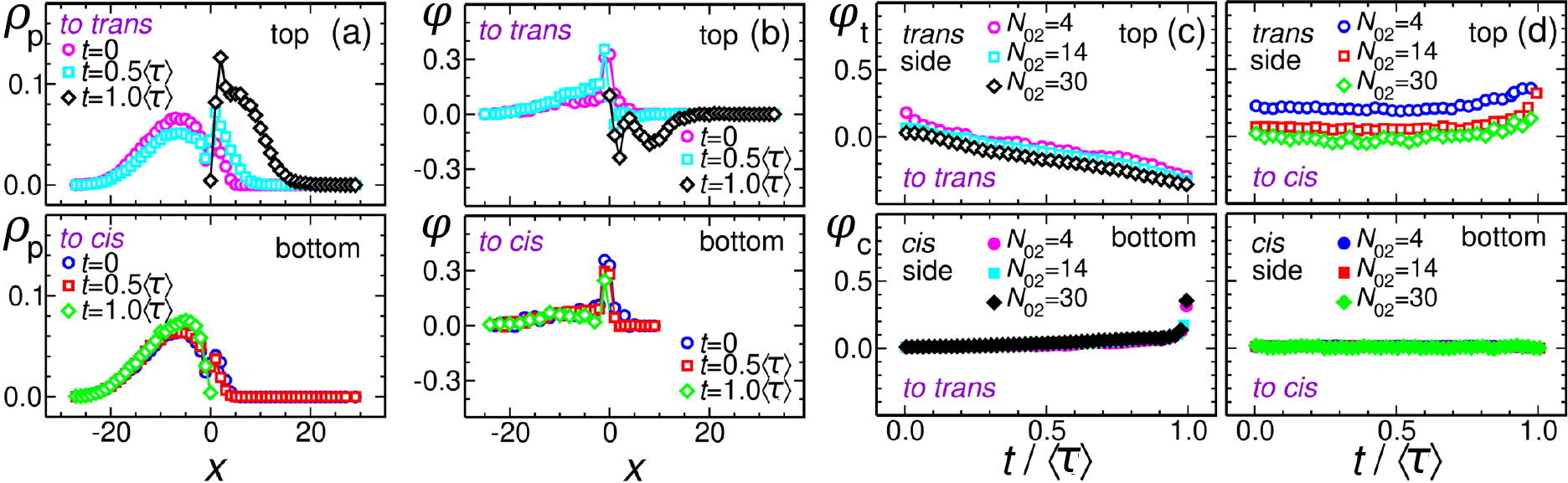}
    \end{center}
    \end{minipage}
\caption{(a) Top and bottom panels show the monomer number density of the polymer chain for the {\it to-trans} and {\it to-cis} states, respectively, as a function of the distance from the nanopore $x$ (pore is at $x=0$) at times $t=0$ (top: open pink circles, bottom: open blue circles), $t=0.5 \langle \tau \rangle $ (top: open turquoise squares, bottom: open red squares) and $t=1.0 \langle \tau \rangle $ (top: open black diamonds, bottom: open green diamonds), with fixed values of $L_{\textrm{r}}=16$, $\rho_{\textrm{r}}=0.04$ ($N_{\textrm{r}}=100$), $F_{\textrm{sp}}=4$ and $N_{02}=N_{02\textrm{m}}=14$ for all panels.
(b) Top: Bond-orientation order parameter $\phi(x)$ for the {\it to-trans} state as a function of the distance to the pore $x$ at  $t=0$ (open pink circles), $t=0.5 \langle \tau \rangle $ (open turquoise squares) and $t=1.0 \langle \tau \rangle $ (open black diamonds). Bottom: The same as top panel but for the {\it to-cis} state.
(c) Top: Bond-orientation order parameter in the {\it trans} side $\phi_{\textrm{t}}$ as a function of normalized time $t /  \langle \tau \rangle $, for the {\it to-trans} state with initial values of the polymer subchain contour length in the {\it trans} side $N_{02} = 4$ (open pink circles), $14$ (open turquoise squares) and $30$ (open black diamonds). Bottom: Order parameter on the {\it cis} side $\phi_{\textrm{c}}$ as a function of normalized time $t / \langle \tau \rangle $, for the {\it to-trans} state with initial values of the polymer subchain contour length in the {\it trans} side $N_{02} = 4$ (filled pink circles), $14$ (filled turquoise squares) and $30$ (filled black diamonds).
(d) The same as panel (c) but for the {\it to-cis} state. 
} 
\label{orientation}
\end{center}
\end{figure*}

\subsection{Waiting time distribution} \label{wtime}

An important quantity which shows the translocation dynamics at the monomer level is the waiting time (WT) distribution, that is the time that each bead spends at the nanopore during the course of translocation process. In order to present the difference between the {\it to-cis} and {\it to-trans} states, in panels (c), (d) and (e) in Fig.~\ref{ttimes} the WT distribution $w$ has been plotted as a function of $(s-N_{02}) / N_{01}$, where $s$ is the monomer index minus one, for fixed values of the rod length, SP force and rod density $L_{\textrm{r}}=16$, $F_{\textrm{sp}}=4$ and {$\rho_{\textrm{r}}=0.04$ (corresponding to $N_{\textrm{r}}=100$)}, respectively. 

Panel (c) in Fig.~\!\ref{ttimes} shows the WT distribution ($w$) of the {\it to-cis} state for the initial {\it tans}-side polymer subchain contour lengths $N_{02} = 4$ (blue line - circles) and $14$ (red line - squares). Obviously, increasing $N_{02}$ slows down the dynamics. 
Panel (d) is the same as (c) but for the {\it to-trans} state for $N_{02} = 4$ (blue line - circles), $14$ (red line - squares) and $30$ (green line - diamonds). Here, the waiting time decreases
as it becomes more probable to form a larger fold for increasing $N_{02}$.
Finally, panel (e) compares the WT distributions for the {\it to-cis} (blue line - circles) and {\it to-trans} (red line - squares) states when the value of the initial {\it trans}-side polymer subchain contour length lies on the border curve in the phase diagram in Fig.~\!\ref{probability}(b) with $N_{02} = N_{02,\rm m} = 14$ for $\rho_{\textrm{r}}=0.04$. 

\subsection{The polymer structure} \label{structure}

To further clarify the role of the folds in the {\it trans} side that assist translocation {\it to-trans}, we have performed a more detailed analysis of the polymer configurations during translocation. In the top panel of Fig.~\!\ref{orientation}(a) we plot the monomer number density profile of the polymer chain $\rho_{\textrm{p}}$ as a function of the distance from the nanopore $x$ at different times $t=0$ (pink circles), $0.5 \langle \tau \rangle $ (turquoise squares) and $1.0 \langle \tau \rangle $ (black diamonds) for the {\it to-trans} state. The {\it cis} and the {\it trans} sides are at $x<0$ and $x>0$, respectively, and the nanopore at $x = 0$. The bottom panel of (a) shows the same data for the {\it to-cis} state. The crowding of the monomers close to the pore in the {\it trans} side indicates that the polymer has folds for the {\it to-trans} state.
To quantify this, we define the bond-orientation order parameter with respect to the membrane as 
\begin{equation}
\phi (x_i) = \bigg\langle \frac{1}{n_i (t)} \sum_{k=1}^{n_i (t)} \big[ 1 - 2 \big( \hat{\textrm{e}}_{i,k} \cdot \hat{u}_{y} \big)^2 \big] \bigg\rangle,
\label{polarity}
\end{equation}
where $x_i$ is the horizontal coordinate of a bin (a thin slab parallel to the wall with thickness of unity), $n_i (t)$ is the number of polymer bonds located in the $i^{\textrm{th}}$ bin at time $t$, $\hat{\textrm{e}}_{i,k}$ is the unit vector of the $k^{\textrm{th}}$ bond in the $i^{\textrm{th}}$ bin, $\hat{u}_y$ is a unit vector parallel to the membrane ($y$ axis), and $\langle \cdots \rangle$ denotes an ensemble average. This quantity measures the average orientation of the bond vectors with respect to the wall. The value of $\phi$ varies between $[-1, +1]$ and its value is $1$, $-1$, or $0$ if the bond orientations are perpendicular, parallel, or randomly distributed with respect to the membrane, respectively.
 
In the top panel of Fig.~\!\ref{orientation}(b) we plot the order parameter $\phi(x)$ for the {\it to-trans} state as a function of $x=x_i$, at different moments $t=0$ (pink circles), $t=0.5 \langle \tau \rangle $ (turquoise squares) and $t=1.0 \langle \tau \rangle $ (black diamonds). With the passage of the time a spatially oscillating profile forms on the {\it trans} side, indicating a  folded structure. In the bottom panel of (b) for the {\it to-cis} state the profile remains mostly flat.

Another useful way to quantify the configurations is to divide the order parameter into two spatially averaged components that evolve in time. They are defined as
\begin{eqnarray}
\phi_{\textrm{t}} (t) &=& \bigg\langle \frac{1}{n_{\textrm{t}}  (t)} \sum_{k=1}^{n_{\textrm{t}}  (t)} \big[ 1 - 2 \big( \hat{\textrm{e}}_{k} \cdot \hat{u}_y \big)^2 \big] \bigg\rangle ; 
\nonumber\\
\phi_{\textrm{c}}  (t)  &=& \bigg\langle \frac{1}{n_{\textrm{c}}  (t)} \sum_{k=1}^{n_{\textrm{c}}  (t)} \big[ 1 - 2 \big( \hat{\textrm{e}}_{k} \cdot \hat{u}_y \big)^2 \big] \bigg\rangle ,
\label{tot_polarity}
\end{eqnarray}
respectively, where $\hat{\textrm{e}}_{k}$ is the unit bond vector for the $k^{\textrm{th}}$ bond either in the {\it trans} side in $\phi_{\textrm{t}} (t)$, or in the {\it cis} side in $\phi_{\textrm{c}} (t)$, and $n_{\textrm{t}}  (t)$ and $n_{\textrm{c}}  (t)$ are the total number of bonds in the {\it trans} and in the {\it cis} sides, respectively, at time $t$. 

These {\it trans}- and {\it cis}-side components are plotted in Figs.~\!\ref{orientation}(c) and (d) for {\it to-trans} and {\it to-cis} states, respectively, as a function of the normalized time $t / \langle \tau \rangle $.
The negative values of $\phi_{\textrm{t}} (t)$ as time progresses indicate folds in the {\it trans} side for the {\it to-trans} state. At the same time tension propagates in the {\it cis} side as shown in more detail in the SI, making $\phi_{\textrm{c}} (t)$ positive because the bonds tend to align perpendicular to the wall.  
On the other hand, for the {\it to-cis} state, in the top panel of (d), the {\it trans}-side polymer subchain retracts to the {\it cis} side and
 $\phi_{\textrm{t}} (t)$ remains positive.  Moreover, as can be seen in the bottom panel of (d), for this case the process is very slow and the {\it cis}-side subchain remains in a state close to equilibrium with $\phi_{\textrm{c}} (t) \approx 0$.

\section{Summary and conclusions} \label{CONCLUSION}

In the present work we have quantified the role of different physical parameters for active-rod-assisted polymer translocation through a nanopore. In this setup the polymer has a fixed number of $N_{01}$ monomers on the {\it cis} side of the membrane and a varying number density of active rods (ARs) $\rho_{\rm r}$ in the {\it trans} compartment. The interaction between ARs and the polymer in the {\it trans} side of the subchain with $N_{02}$ initial monomers facilitates successful {\it to-trans} events even for $N_{02} < N_{01}$, which is not possible for non-driven translocation. We map out state diagram of successful events in the $N_{02,\rm m} - \rho_{\rm r}$ plane revealing competition between facilitation and crowding by the ARs, which results in an optimal value of $\rho_{\rm r}$ for a mimimum value of $N_{02,\rm m}$ and correspondingly to fastest translocation. We have further done detailed analysis of the configurations of the translocating polymers which shows that successful translocation to the {\it to-trans} state is accompanied by formation of folds in the {\it trans} side and crowding of the rods near the membrane wall. These results may shed light on better understanding polymer  translocation process into or out of the living cells which typically contain various kinds of active particles.  


\begin{acknowledgments}
Computational resources from CSC - Center for Scientific Computing Ltd. and 
from the Aalto University School of Science “Science-IT” project are gratefully acknowledged.
T.A-N. has been supported in part by the Academy of Finland through its 
PolyDyna (no. 307806) and QFT Center of Excellence Program grants (no. 312298).
\end{acknowledgments}


\end{document}